%% file: main.tex
\documentclass{./sig-alternate}

\usepackage{mathptmx}
\usepackage{subcaption}
\usepackage{fancyhdr}
\usepackage[normalem]{ulem}
\usepackage[hyphens]{url}
\usepackage[sort,nocompress]{cite}
\usepackage[final]{microtype}
\usepackage[keeplastbox]{flushend}
\usepackage[inline]{enumitem}
\usepackage[draft=true]{minted}

\usepackage[hidelinks]{hyperref}
\fancypagestyle{firstpage}{
  \fancyhf{}
  
  \fancyhead[C]{\vspace{10pt}\normalsize{MICRO 2022 Submission
      \textbf{\#\microsubmissionnumber} -- Confidential Draft -- Do NOT Distribute!!}\\\vspace{-25pt}} 
  \fancyfoot[C]{\thepage}
}

\title{Canal: A Flexible Interconnect Generator for Coarse-Grained Reconfigurable Arrays} 

\author{Jackson Melchert*, Keyi Zhang*, Yuchen Mei, Mark Horowitz, Christopher Torng, Priyanka Raina\\Stanford University; *Equal Contribution}

\begin{document}
\maketitle
\pagestyle{plain}

\input{sections/1-abstract}
\input{sections/2-introduction}

\input{sections/3-system-design}

\input{sections/4-evaluation}

\input{sections/5-conclusion}

\bibliographystyle{IEEEtranS}
\bibliography{refs}

\end{document}

%% file: sections/1-abstract.tex
\begin{abstract}

The architecture of a coarse-grained reconfigurable array (CGRA) interconnect has a significant effect on not only the flexibility of the resulting accelerator, but also its power, performance, and area. Design decisions that have complex trade-offs need to be explored to maintain efficiency and performance across a variety of evolving applications. This paper presents Canal, a Python-embedded domain-specific language (eDSL) and compiler for specifying and generating reconfigurable interconnects for CGRAs. Canal uses a graph-based intermediate representation (IR) that allows for easy hardware generation and tight integration with place and route tools. We evaluate Canal by constructing both a fully static interconnect and a hybrid interconnect with ready-valid signaling, and by conducting design space exploration of the interconnect architecture by modifying the switch box topology, the number of routing tracks, and the interconnect tile connections. Through the use of a graph-based IR for CGRA interconnects, the eDSL, and the interconnect generation system, Canal enables fast design space exploration and creation of CGRA interconnects.

\end{abstract}

%% file: sections/2-introduction.tex
\section{Introduction}
\vspace{-0.4cm}

Coarse-grained reconfigurable arrays (CGRAs) have been studied heavily in recent years as a promising configurable accelerator architecture~\cite{adres, dyser, plasticine, artem}. The end of Moore's law necessitates the creation of specialized hardware accelerators to enable running increasingly complex image processing and machine learning applications. While a variety of hardware accelerator architectures exist, CGRAs have emerged as an interesting midpoint between the flexibility of an FPGA and the performance of an application-specific accelerator. A CGRA can achieve high energy efficiency and performance due to word-level arithmetic operations and interconnect, while maintaining enough flexibility to run a variety of applications that evolve over time \cite{ambervlsi}.

\begin{figure}
    \centering
    \includegraphics[width=0.33\textwidth]{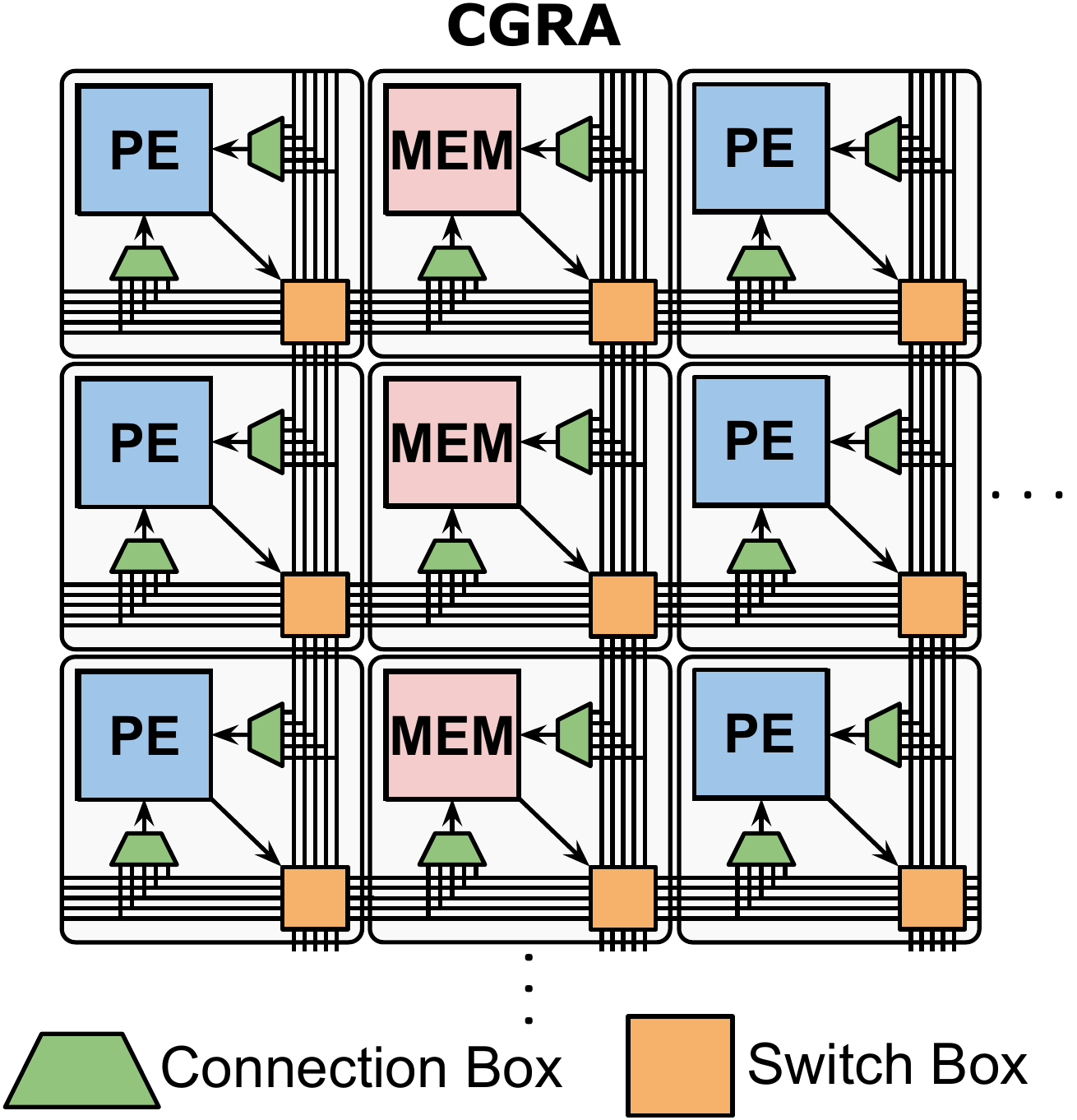} 
    \caption{Architecture of a CGRA with PE tiles, memory tiles, connection boxes, and switch boxes.}
    \label{fig:cgra-diagram}
\end{figure}

CGRAs, as well as other spatial accelerator architectures, often have hundreds of compute cores and memory cores. These compute cores (called processing elements or PEs) and memory (MEM) cores are laid out spatially in a grid of tiles and are connected through a configurable interconnect. An example is shown in Fig.~\ref{fig:cgra-diagram}. The reconfigurable interconnect contains switch boxes (SBs), which connect the PE/MEM outputs to the tracks in the interconnect, and connection boxes (CBs), which connect the interconnect tracks to the inputs of the cores. While having a large number of compute cores enables very high performance, the reconfigurable interconnect connecting these cores can constitute over 50\% of the CGRA area and 25\% of the CGRA energy \cite{artem}. Design space exploration of the interconnect is necessary to achieve high performance with lower energy and area costs. 

There are many interconnect design choices that directly impact the power, performance, and area of the resulting accelerator, including the number and bitwidth of tracks in the interconnect, how the processing elements and memories are arranged in the array, how those elements are connected, and how the interconnect is configured. An agile approach for specifying and generating the interconnect is needed for efficient design space exploration.

In this paper, we present Canal, a Python-embedded domain-specific language (eDSL) and compiler for specifying and generating reconfigurable interconnects for CGRAs using a graph-based intermediate representation. The major contributions of our paper are:
\begin{enumerate}
    \item We describe a graph-based intermediate representation (IR) for CGRA interconnects that is capable of representing and  generating a variety of topologies.
    \item We propose an embedded domain-specific language (eDSL) called Canal that can compile an interconnect architecture specification into the graph-based IR.
    \item We propose an interconnect generator system that can take the IR and automatically produce hardware, place and route collateral, and a bitstream generator.
    \item We explore various design space choices using Canal and demonstrate its effectiveness in generating an efficient CGRA design.
\end{enumerate}

\section{Related Work}
\vspace{-0.4cm}
Previous attempts have been made to create an interconnect generator for CGRAs, but new demands in CGRA design necessitate a more flexible and powerful system. VPR is one of the most established FPGA architecture research tools~\cite{vpr}. It allows users to adjust various design aspects of the FPGA and observe the effects on final application performance, such as timing and area usage. However, VPR does not offer an RTL generator and users have to design their FPGA independently and hand-write the VPR architecture file accordingly. 

CGRA-ME~\cite{cgra-me} is a CGRA architecture research tool similar to Canal in that it also offers integrated RTL generation and place and route tools. One of the major differences is the architecture specification. CGRA-ME opts for a more rigid XML-based input whereas Canal takes in a Python eDSL program, which is more flexible and readable. 

FastCGRA~\cite{fastcgra} is a similar CGRA architecture exploration tool that uses an eDSL to construct the hardware. However, with FastCGRA, users explicitly construct multiplexers and switches, from which RTL is generated. Canal abstracts away the notion of hardware primitives and lets the compiler backend choose how to generate the hardware. In summary, Canal supports more tools, allows for more flexibility, and enables easier design space exploration than previous attempts at an interconnect generator.


%% file: sections/3-system-design.tex
\section{System Design}
\begin{figure}
    \centering
    \includegraphics[width=0.48\textwidth]{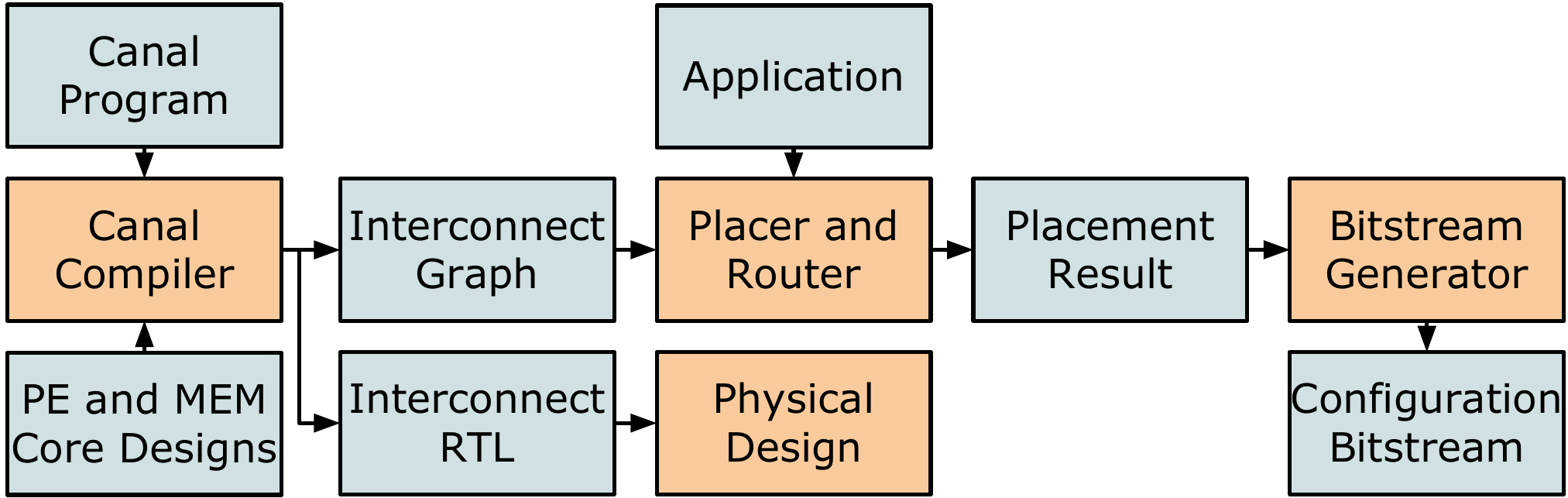} 
    \caption{The Canal interconnect generator system. It takes an interconnect specification written as a program in the Canal eDSL and produces the interconnect RTL implementation. Canal also takes an application and places and routes it on a CGRA with the specified interconnect and generates a configuration bitstream.}
    \label{fig:system-diagram}
    \vspace{-1.4em}
\end{figure}

\vspace{-0.4cm}
In this section, we introduce the design of the Canal system, including the graph-based IR for representing interconnects, the Canal eDSL, the static interconnect generation used to translate the IR into hardware, and finally how that IR interfaces with the application place and route (PnR) algorithms. Fig.~\ref{fig:system-diagram} summarizes how the Canal interconnect generator interfaces with the PE and memory core designs, application PnR, RTL generation, and bitstream generation. 

\subsection{Graph-Based Intermediate Representation}
\begin{figure}
    \centering
    \captionsetup{justification=centering}
    \begin{subfigure}[b]{.2\textwidth}
        \centering
        \includegraphics[width=0.8\textwidth]{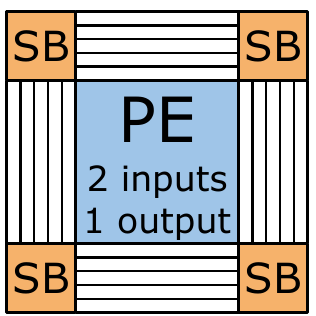} 
        \caption{Hardware representation of a PE with four SBs}
    \end{subfigure}
    \hfill
    \begin{subfigure}[b]{.2\textwidth}
        \centering
        \includegraphics[width=0.8\textwidth]{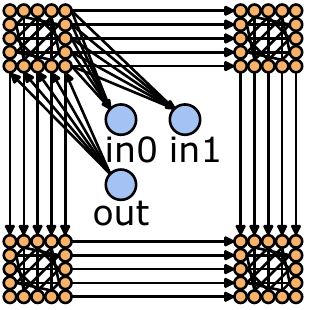} 
        \caption{Digraph representation of a PE with four SBs}
    \end{subfigure}
    \captionsetup{justification=justified}
    \caption{Hardware and directed graph based intermediate representation of the configurable interconnect. Not all connections between the PE and SBs are shown for simplicity.}
    \label{fig:graph-representation}
    \vspace{-0.5em}
\end{figure}

\vspace{-0.4cm}
The primitives in Canal's intermediate representation (IR) are nodes, which represent anything that can be connected in the underlying hardware, and edges, which are wires connecting the nodes together. An example of the IR for a switch box is shown in Fig.~\ref{fig:graph-representation}.

All edges are unidirectional so the IR represents a directed graph. Nodes in the graph can have multiple incoming edges which, when translated into hardware, transform into multiplexers. Each node also has attributes that provide additional information for type checking and hardware generation.

This intermediate representation is flexible enough to represent a wide variety of interconnect topologies and handle an arbitrarily complex set of CGRA cores. We will discuss a few design space exploration experiments that exploit this flexibility in Section~\ref{sec:results}.



\subsection{The Canal Language}
\vspace{-0.4cm}
\begin{figure}[t]
    \centering

\begin{minted}{python}
node = Node(x=1, y=1, side="south", track=1)
for port_node in tile.pe.inputs():
    node.add_edge(port_node)
    
create_uniform_interconnect(width=32, 
    height=32, sb_type="wilton", num_tracks=5, 
    track_width=16, reg_density=1)
\end{minted}
\vspace{-1em}

    \caption{Example Canal low level node creation and high level interconnection creation. Canal includes many useful high level interconnect construction functions to enable the creation of common interconnects.}
    \label{fig:canal_example}
    \vspace{-1.3em}
\end{figure}

The Canal language is a Python-embedded domain-specific language (eDSL) that constructs the interconnect intermediate representation described in the previous section. The Canal language translates the Python description of an interconnect into this IR, so at the lowest level, a designer could instantiate nodes in the Canal language and wire them together.

As Canal is embedded in Python, we have also built a layer on top of the basic primitives of Canal, which simplifies the IR construction. For instance, for creating a uniform interconnect (all switch boxes have the same topology) with no diagonal connections, we provide a simple helper function that produces different interconnect topologies by varying function parameters such as height and width of the array, switch box topology, number of tracks, bit width of tracks, and density of pipeline registers. An example of creating a low level Canal node and using a higher level Canal helper function is shown in Fig.~\ref{fig:canal_example}.

The Canal eDSL allows designers to easily conduct design space exploration by varying the parameters in the helper functions, or by generating an entirely new interconnect. Canal can easily be integrated with other DSLs. For example, one could use a DSL for specifying individual CGRA tiles and then integrate them together using Canal to generate the interconnect.

\subsection{Generating Interconnect Hardware}
\label{sec:compile-hardware}
\vspace{-0.4cm}
Because Canal's IR only describes the connectivity among the different nodes, it is up to the hardware compiler backend to decide how to lower the IR. We implemented two different hardware compiler backends that lower the IR into \begin{enumerate*}[label=(\arabic*)] \item a static mesh interconnect and \item a statically configured network-on-chip (NoC) \end{enumerate*}. This NoC has data channels along with a ready-valid interface and routing is configured statically before the application runs. We use magma~\cite{magma} as our hardware circuit implementation, but this could be extended to any hardware generator framework.

To generate a static mesh interconnect, we adopt the following principles to generate hardware:

\begin{enumerate}
    \item Nodes with hardware attributes (e.g. a processing element core) generate the specified hardware.
    \item Directed edges are translated into wires.
    \item Nodes with multiple incoming edges generate multiplexers.
\end{enumerate}

\begin{figure}
    \centering
    \includegraphics[width=0.43\textwidth]{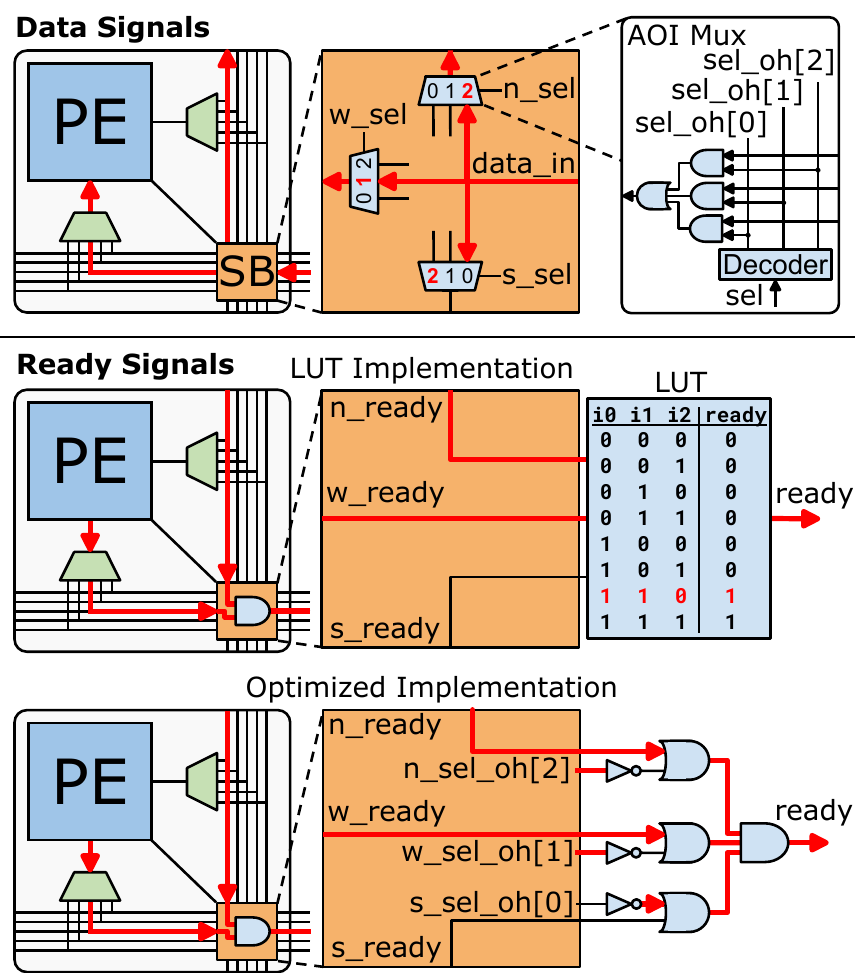} 
    \caption{Joining logic for the ready signals on the configurable interconnect. In this example, the data signal coming in from the east is being routed to both the north and west. The ready signals corresponding to the data signals flow in the opposite direction through the same switch boxes. The two ready signals need to be joined together to form the final ready signal, and we can reuse the one-hot switch box mux select signals to compute this.}
    \label{fig:ready-valid-join}
    \vspace{-1em}
\end{figure}

\begin{figure}
    \centering
    \includegraphics[width=0.25\textwidth]{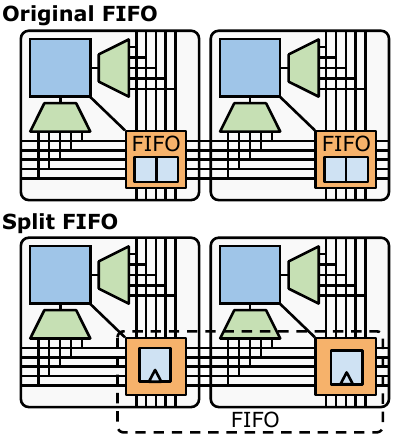} 
    \caption{Split FIFO optimization where two registers in adjacent switch boxes can function as a FIFO.}
    \label{fig:split-fifo}
    \vspace{-2em}
\end{figure}

We also use attributes associated with each node to lower the node to different hardware components. For instance, a register node will be lowered into a physical register. A port node will be lowered to a CB (with an internal multiplexer) where the output of the CB connects to the port of the core. These translations are mechanical and can be accomplished through a compiler pass.

Reusing the same IR to generate a statically configured NoC has several challenges. 
First, the application graph may have fanouts, that is, one output port of a node is connected to multiple input ports. While this is simple to handle in a static interconnect, we now need an area-efficient way to handle control signals.

Since valid signals flow in the same direction as the data, generating hardware for valid channels follows the same strategy as that used for the data channels. However, since ready signals flow in a direction opposite to that of the data channels, we need a way to merge ready signals at the fan-in point. A naive solution is to implement a lookup table (LUT) that encodes the fan-in information, as shown in Fig.~\ref{fig:ready-valid-join}. During configuration, we statically encode the ready signal joining logic into the LUT. However, building a LUT for each multiplexer is expensive and it also bloats the configuration space. To optimize this logic, we leverage the fact that the data multiplexers in the switch box are AOI multiplexers, which internally use a decoder to convert the mux selection bits into a one-hot vector, which represents the active routing information. The ready signal joining logic needs this routing information to properly join multiple ready signals. We can reuse these one-hot decoder signals to compute the joining logic without introducing expensive LUT based solutions, as shown at the bottom of Fig.~\ref{fig:ready-valid-join}. In this figure, the one-hot selection signals are $n\_sel\_oh$, $w\_sel\_oh$, and $s\_sel\_oh$. We know that if bit 2 of $n\_sel\_oh$ is high, than the $data\_in$ is routed to the north. We OR the inversion of $n\_sel\_oh[2]$ with the $n\_ready$ signal, giving us a signal that is high when $n\_ready$ is high or when that route is not used. We can repeat this logic for the remaining two directions and AND them all together to form the final $ready$ output.

Another challenge when generating a statically configured NoC is that a ready-valid NoC needs FIFOs present in the interconnect to buffer data when a downstream tile is not yet ready. While we can easily generate a fixed-size FIFO for a register IR node, the area cost of those FIFOs can be quite high, as shown in Fig~\ref{fig:sb-fifo-area}. Therefore, we also need a way to reduce FIFO size.
To reduce the area overhead introduced by FIFOs while maintaining backward compatibility with a static interconnect, we realize that we can combine two registers from adjacent tiles into a single size-two FIFO. We call this a split FIFO. The first register's FIFO control signals are passed from the first tile into the second tile and its register, as shown in Figure~\ref{fig:split-fifo}.

We can also chain more registers together into a deeper FIFO using the same logic. Since each register's FIFO logic is slightly different depending on its location inside the FIFO pipeline, we need to configure them differently during place and route. The drawback, however, is that these control signals cannot be registered at the tile boundary; the longer the FIFO is chained, the longer the combinational delay on the path. However, if most of the target applications do not require a deep FIFO between different nodes, using this scheme can significantly reduce the silicon area. The area impact of this optimization is evaluated in Fig.~\ref{fig:sb-fifo-area}.


\begin{figure}
    \centering
    \captionsetup{justification=centering}
    \begin{subfigure}[b]{.22\textwidth}
        \includegraphics[width=1\textwidth]{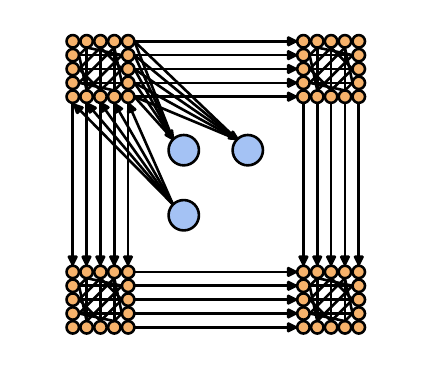} 
        \caption{Digraph representation with no edge weights}
    \end{subfigure}%
    \begin{subfigure}[b]{.26\textwidth}
        \includegraphics[width=0.8\textwidth]{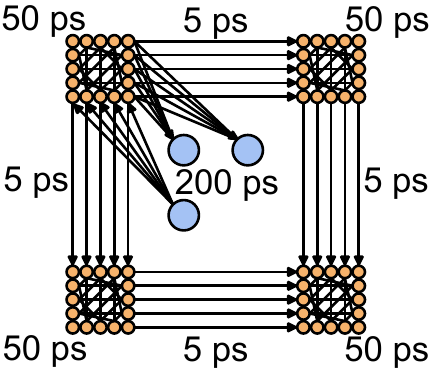} 
        \caption{Digraph representation with timing information as weights}
    \end{subfigure}
    \captionsetup{justification=justified}
    \caption{The edge weights of the directed graph representation of the interconnect allow the PnR algorithm to be run on the graph directly.}
    \label{fig:pnr}
    
\end{figure}

After the graph is translated into RTL, Canal verifies structural correctness by comparing the connectivity of the hardware with that of the IR by parsing the generated RTL. In addition, Canal also has a built in configuration sweep test suite that exhaustively tests every possible connection in IR on the CGRA. This ensures correctness of the design. 

The methodology described here also applies to generating dynamic NoCs. Instead of lowering a node into a configurable multiplexer to select among incoming data tracks, we can generate a router whose routing table is computed based on the same connectivity information.

\subsection{Place and Route using Canal Interconnect}
\vspace{-0.4cm}

This section describes how the Canal system integrates with a PnR algorithm backend (see Fig.~\ref{fig:system-diagram}) to enable running applications on a given interconnect. During the translation from a Canal program into the directed graph representation, information regarding important hardware characteristics, like core or wire delays, can be embedded into the graph as shown in Fig.~\ref{fig:pnr}. The Canal system then executes PnR in three stages: packing, placement, and routing. The remainder of this section describes the PnR backend we use in our results.

During the packing stage, both the interconnect graph and the application graph (represented as a dataflow graph) are loaded into the PnR tool. Constants and registers in the application are analyzed to identify any packing opportunities. For example, a pipeline register that feeds directly into a PE can be packed within that PE, eliminating the need to place that register on the configurable interconnect. 

After packing, the placement tool places the tiles in the application onto the interconnect in two stages: global placement and detailed placement. 
Global placement uses an analytical algorithm that leverages the standard conjugate gradient method on the summation of the cost of each net (Equation~\ref{eqn:globalplacement-function})~\cite{kahng2005aplace}. The cost of a net is the combination of its half-perimeter wire length (HPWL) and a legalization term for memory tiles. In global placement, we use L2 distance to approximate the HPWL to speed up the algorithm. CGRAs typically have fewer rows or columns of memory tiles (compared to PE tiles), so the legalization term is needed to ensure that memory tiles are only placed in those rows/columns.
\vspace{-0.2em}
\begin{equation}
\label{eqn:globalplacement-function}
\text{Cost}_{net} = \text{HPWL}_{net, estimate} + \text{MEM}_{potential}
\end{equation}

After global placement, we perform detailed placement based on simulated annealing~\cite{van1987simulated}. The cost function for simulated annealing (see Equation~\ref{eqn:detailedplacement-function}) is the total wirelength of the application, calculated by summing the HPWL cost for each net, and an additional term for penalizing pass-through tiles. Pass-through tiles are those that are only used for routing, which need to be powered on despite not computing anything for the application. $\gamma$ and $\alpha$ are hyperparameters. Some CGRAs have tile-level power gating that enables the ability to turn off tiles that are used for neither the application's computation nor as pass-through tiles. A higher value of $\gamma$ penalizes pass-through tiles more, which encourages the placement algorithm to use already-used tiles for routing, rather than powering on otherwise unused tiles. A higher value for $\alpha$ will penalize longer potential routes, thereby encouraging shorter critical paths after routing. We find that sweeping $\alpha$ from 1 to 20 and choosing the best result post-routing results in short application critical paths.
\vspace{-0.2em}
\begin{equation}
\label{eqn:detailedplacement-function}
\text{Cost}_{net} = (\text{HPWL}_{net} - \gamma \times (\text{Area}_{net} \cap \text{Area}_{existing}))^\alpha
\end{equation}

After global and detailed placement, we route using an iteration-based routing algorithm~\cite{swartz1998fast}. During each iteration, we compute the slack on a net and determine how critical it is given global timing information. Then we route using the A* algorithm on the weighted graph. The weights for each edge are based on historical usage, net slack, and current congestion. This allows us to balance both routing congestion and timing criticality. Similarly to detailed placement, we also adjust the wire cost functions to discourage the use of unused tiles in favor of tracks within already-used tiles. We finish routing when a legal routing result is produced.

%% file: sections/4-evaluation.tex
\begin{figure}
    \centering
    \includegraphics[width=0.26\textwidth]{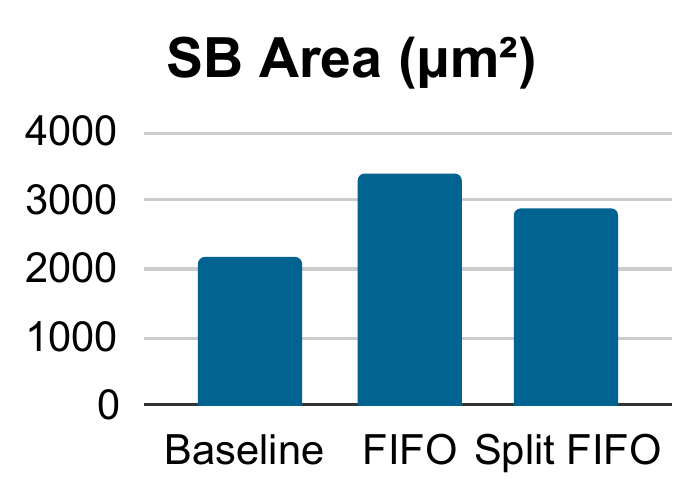} 
    \vspace{-1em}
    \caption{Area comparison of a baseline fully static switch box, a switch box that includes FIFOs for ready/valid applications, and an optimized switch box with a split FIFO.}
    \label{fig:sb-fifo-area}
        \vspace{-1em}
\end{figure}

\begin{figure}
    \centering
    \includegraphics[width=0.3\textwidth]{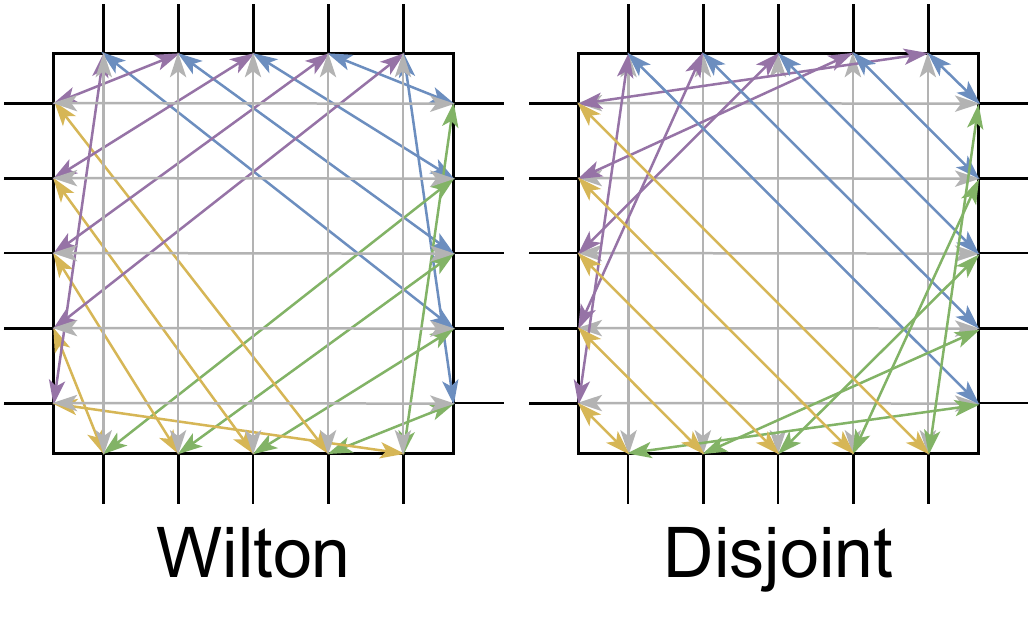} 
    \vspace{-1em}
    \caption{Topology of a Wilton and Disjoint switch box.}
    \label{fig:sb-topos}
    \vspace{-1em}
\end{figure}

\section{Evaluation}
\label{sec:results}
\vspace{-0.4cm}
We evaluate the Canal system by first exploring the optimizations of interconnect FIFOs described in Section~\ref{sec:compile-hardware} and then by using the Canal system to conduct design space exploration of a CGRA interconnect. 

\subsection{Interconnect FIFO Optimizations}
\label{sec:results-fifo}
\vspace{-0.4cm}
We evaluate the effect of introducing FIFOs in the interconnect on switch box area. As described in Section~\ref{sec:compile-hardware}, we need to include FIFOs in the configurable routing when running applications with ready-valid signaling.

As a baseline, we compare against a fully static interconnect with five 16-bit routing tracks containing PEs with two outputs and four inputs, synthesized in Global Foundries 12 nm technology.
As shown in Fig.~\ref{fig:sb-fifo-area}, adding these depth two FIFOs to the baseline design introduces a 54\% area overhead. Splitting the FIFO between multiple switch boxes results in only a 32\% area overhead over the baseline. This optimization allows for much more efficient implementation of an interconnect that supports ready-valid signaling.

\subsection{Interconnect Design Space Exploration}
\vspace{-0.4cm}

We use Canal to explore three important design space axes of a configurable interconnect: switch box topology, number of routing tracks, and number of switch box and connection box port connections.
We find that Canal's automation greatly simplifies the procedure to explore each option in the following subsections.

\begin{figure}
    \centering
    \includegraphics[width=0.23\textwidth]{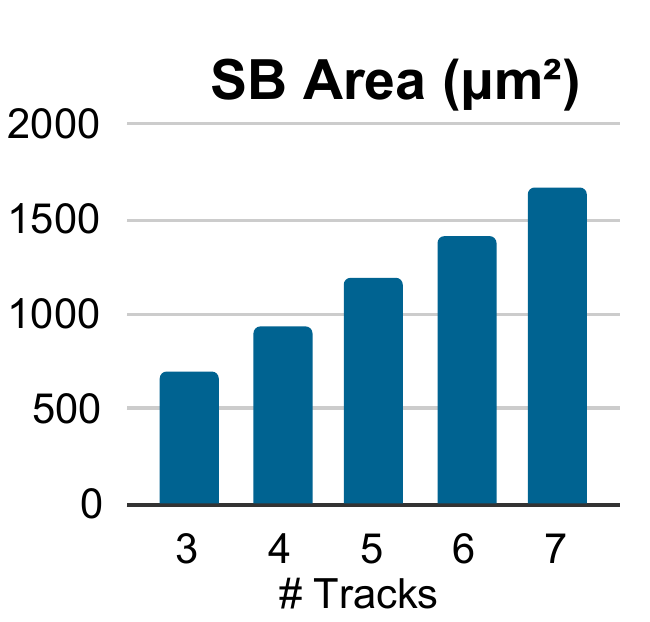} 
    \includegraphics[width=0.23\textwidth]{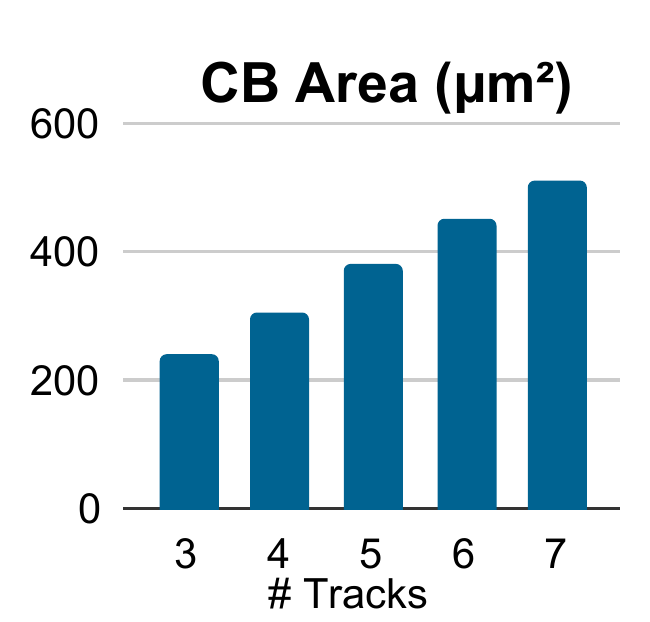} 
    \vspace{-1em}
    \caption{Left: Area of a switch box as the number of tracks increases. Right: Area of a connection box as the number of tracks increases.}
    \label{fig:sb-topo-tracks-area}
    \vspace{-1em}
\end{figure}

\subsubsection{Exploring Switch Box Topologies and Number of Routing Tracks}
\vspace{-0.4cm}

The switch box topology defines how each track on each side of the switch box connects to the tracks on the remaining sides of the switch box. The choice of topology affects how easily nets can be routed on the interconnect. High routability generally corresponds to shorter routes and shorter critical paths in applications. This allows the CGRA to be run at higher frequencies, which decreases application run time. For these experiments we investigate two different switch box topologies illustrated in Fig.~\ref{fig:sb-topos}: Wilton~\cite{wilton} and Disjoint~\cite{disjoint}. These switch box topologies have the same area, as they both connect each input to each of the other sides once.

\begin{figure}
    \centering
    \vspace{-1em}
    \includegraphics[width=0.45\textwidth]{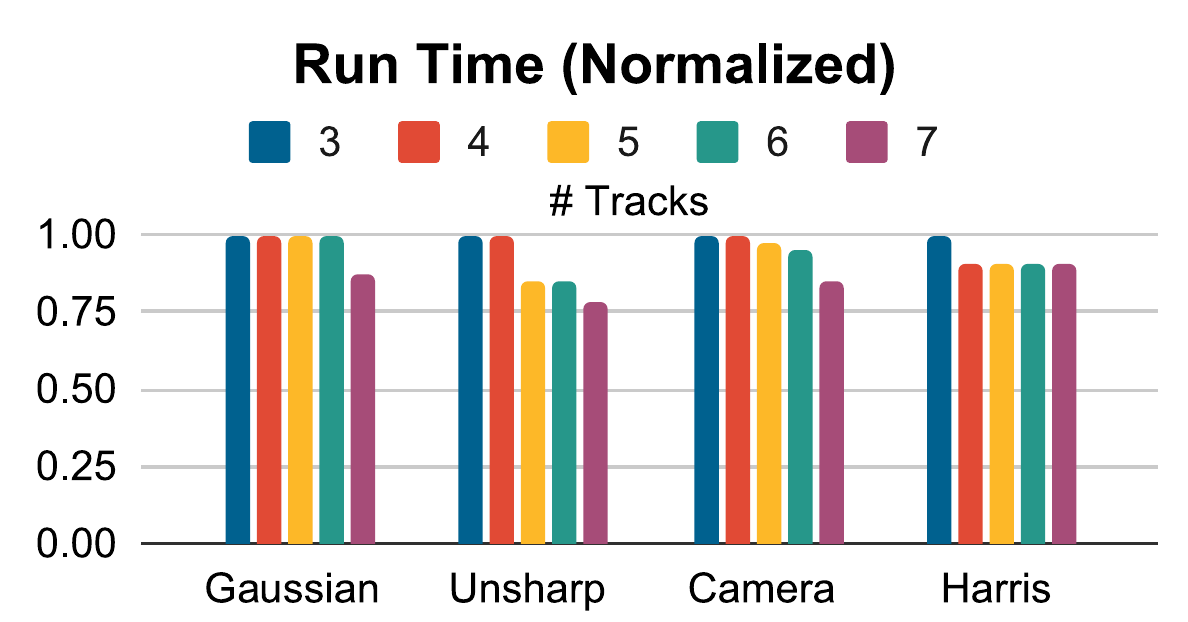}
    \vspace{-1em}
    \caption{Application run time comparison on CGRAs with switch boxes that have different number of tracks.}
    \label{fig:sb-topo-tracks-runtime}
    \vspace{-1em}
\end{figure}

We found that the Wilton topology performs much better than the Disjoint topology, which failed to route in all of our test cases. The Disjoint topology is worse for routability because every incoming connection on track $i$ has a connection only to track $i$ on the three other sides of the SB. This imposes a restriction that if you want to route a wire from any point on the array to any other point on the array starting from a certain track number, you must only use that track number. In comparison, the Wilton topology does not have this restriction resulting in many more choices for the routing algorithm, and therefore much higher routability~\cite{imran}. 

We also vary the number of routing tracks in the interconnect. This directly affects the size of both the connection box and switch box and the amount of routing congestion. For these experiments we measure the area of the connection box and switch box as well as the run time of applications running on the CGRA.

In this experiment we use an interconnect with five 16-bit tracks and PE tiles that have 4 inputs and 2 outputs. As shown in Fig.~\ref{fig:sb-topo-tracks-area}, the area of both the switch box and connection box scale with the number of tracks. From Fig.~\ref{fig:sb-topo-tracks-runtime}, we can see that the run time of the applications generally decreases as the number of tracks increases, although the benefits are less than 25\%. 


\subsubsection{Exploring Switch Box and Connection Box Port Connections}
\vspace{-0.4cm}

\begin{figure}
    \centering
    \vspace{-0.5em}
    \includegraphics[width=0.35\textwidth]{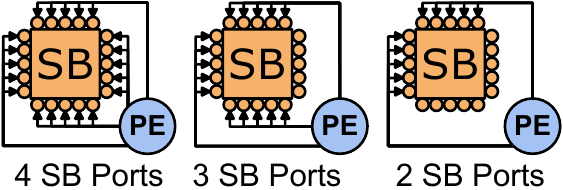} 
    \vspace{-0.5em}
    \caption{Reducing the number of connections from the outputs of the PE to the outgoing ports of the switch box.}
    \label{fig:sb-conns}
    \vspace{-1em}
\end{figure}

Finally, we explore how varying the number of switch box and connection box port connections affects the area of the interconnect and the run time of applications executing on the CGRA. In Canal, we have the ability to specify how many of the incoming tracks from each side of the tile are connected to the inputs/outputs of the PE/MEM cores. Decreasing these connections should reduce the area of the interconnect, but may decrease the number of options that the routing algorithm has. For these experiments, we vary the number of connections from the incoming routing tracks through the connection box to the inputs of the PE/MEM core, and vary the number of connections from the outputs of the core to the outgoing ports of the switch box. At maximum, we can have 4 SB sides, with connections from the core output to the four sides of the switch box. We then decrease this by removing the connections facing east for a total of three sides with connections, and finally we also remove the connections facing south for a total of two sides with connections. This is shown in Fig.~\ref{fig:sb-conns}. We do the same for the connection box.


\begin{figure}
    \centering
    \vspace{-0.2em}
    \captionsetup{justification=centering}
    \begin{subfigure}[b]{.22\textwidth}
        \includegraphics[width=1\textwidth]{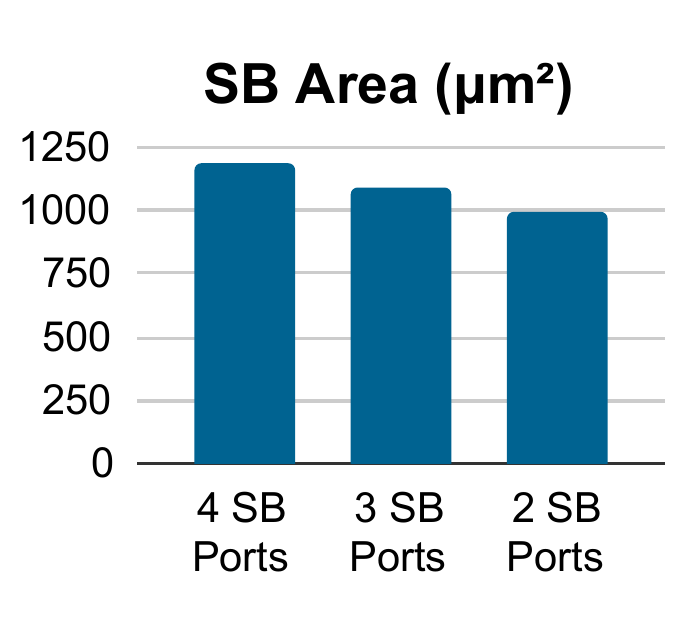} 
    \end{subfigure} \hspace{1em}
    \begin{subfigure}[b]{.22\textwidth}
        \includegraphics[width=0.97\textwidth]{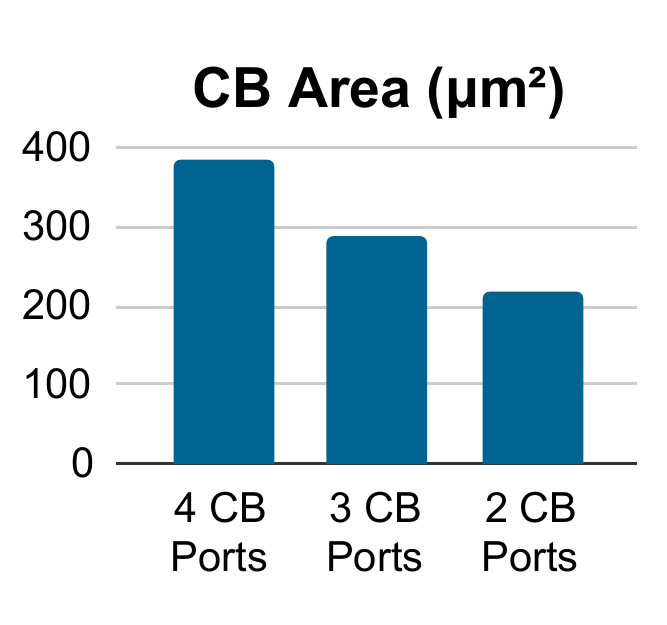} 
    \end{subfigure}
    \vspace{-1em}
    \captionsetup{justification=justified}
    \caption{Area comparison of a switch box and a connection box that have varying number of connections with the four sides of the tile.}
    \label{fig:port-conn-area}
    \vspace{-1em}
\end{figure}

As shown in Fig.~\ref{fig:port-conn-area}, as the number of connections from the core to the switch box decreases, we see a decrease in switch box area. From Fig.~\ref{fig:port-conn-sb-runtime}, we can see that this generally has a small negative effect on the run time of the applications. In this case, a designer could choose to trade some performance for a decrease in switch box area.

As shown in Fig.~\ref{fig:port-conn-area}, as the number of connections from the connection box to the tile inputs decreases, we see a larger decrease in connection box area. From Fig.~\ref{fig:port-conn-cb-runtime}, we can see that this has a larger negative effect on the run time of the applications. Again, a designer could choose to trade some performance for a decrease in connection box area.

%% file: sections/5-conclusion.tex
\section{Conclusion}
\vspace{-0.4cm}

We have developed Canal, a domain-specific language and interconnect generator for coarse-grained reconfigurable arrays. The Canal language allows a designer to easily specify a complex configurable interconnect, while maintaining control over the low-level connections. The hardware generator, placer and router, and bitstream generator help to facilitate design space exploration of CGRA interconnects. We demonstrate the flexibility of Canal by creating a hybrid ready-valid interconnect and demonstrate the design space exploration capabilities of Canal by evaluating different switch box topologies, number of interconnect routing tracks, and number of SB and CB port connections. The power and flexibility that Canal provides will enable more designers to create and explore diverse and interesting CGRA architectures. 

\begin{figure}
    \centering
    \vspace{-1em}
    \includegraphics[width=0.4\textwidth]{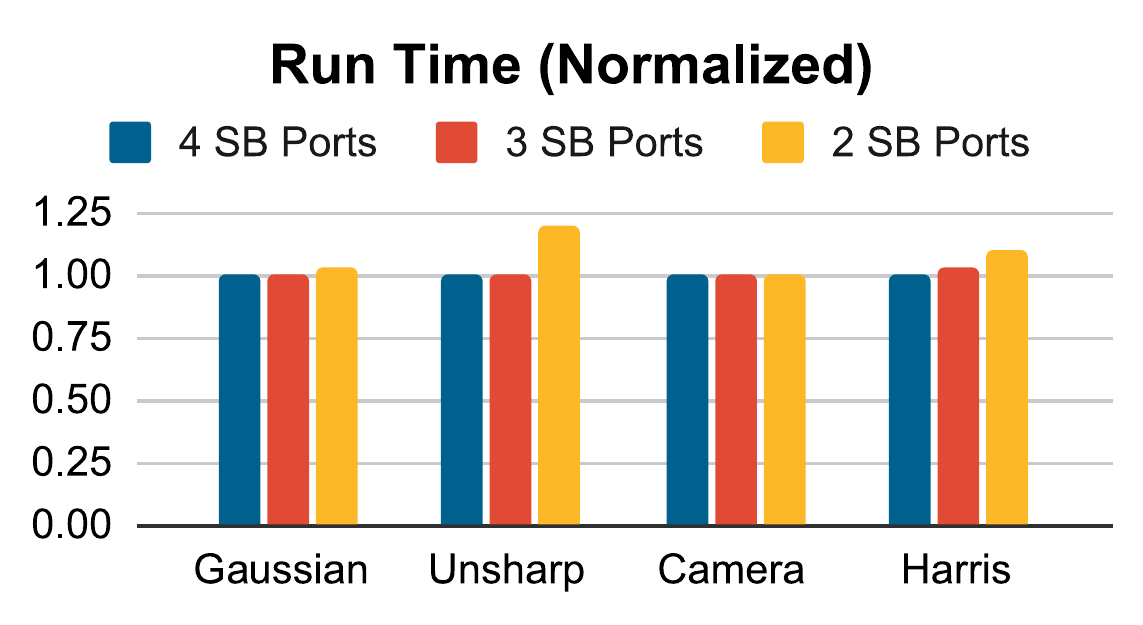} 
    \vspace{-1em}
    \caption{Run time comparison of a switch box that has varying number of connections from the four sides of the tile.}
    \label{fig:port-conn-sb-runtime}
    \vspace{-0.5em}
\end{figure}

\begin{figure}
    \centering
    \vspace{-0.7em}
    \includegraphics[width=0.4\textwidth]{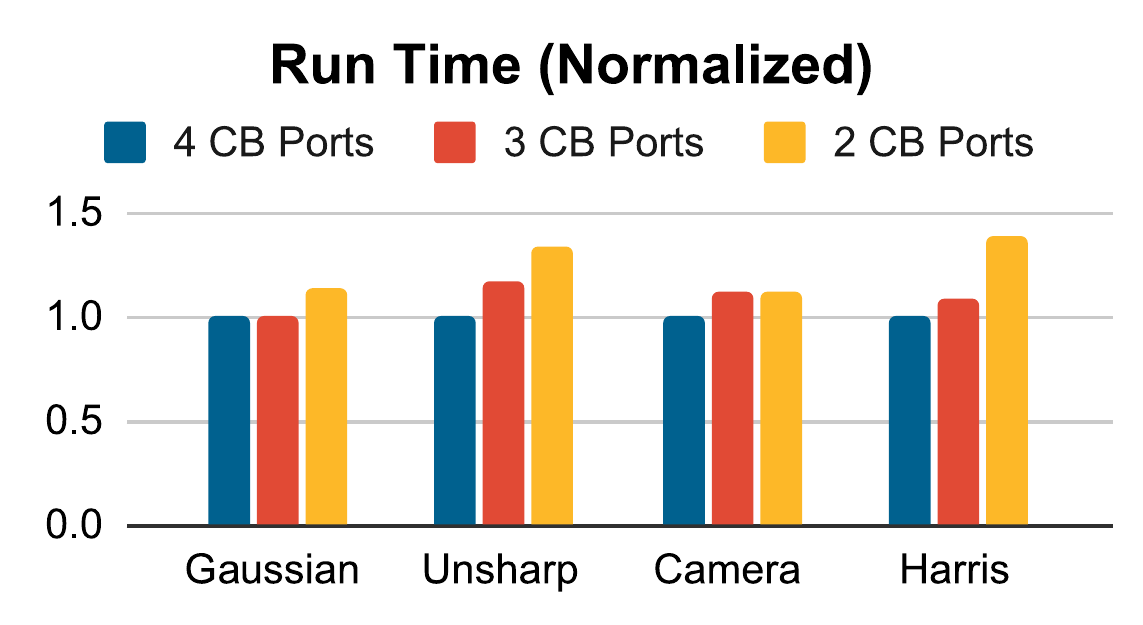} 
    \vspace{-1em}
    \caption{Run time comparison of a connection box that has varying number of connections from the four sides of the tile.}
    \label{fig:port-conn-cb-runtime}
\end{figure}